\documentclass[a4paper]{article}

\usepackage{INTERSPEECH2022}
\usepackage{booktabs}
\usepackage{multirow}
\usepackage{url}
\usepackage[hang,flushmargin]{footmisc}

\title{Pre-trained Speech Representations as Feature Extractors for Speech Quality Assessment in Online Conferencing Applications}
\name{Bastiaan Tamm$^1$, Helena Balabin$^1$, Rik Vandenberghe$^1$, Hugo Van hamme$^2$}
\address{
  $^1$Laboratory for Cognitive Neurology, KU Leuven, Belgium\\
  $^2$Processing Speech and Images, KU Leuven, Belgium}
\email{bastiaan.tamm@kuleuven.be}

\begin{document}

\maketitle
\begin{abstract}
Speech quality in online conferencing applications is typically assessed through human judgements in the form of the mean opinion score (MOS) metric.
Since such a labor-intensive approach is not feasible for large-scale speech quality assessments in most settings, the focus has shifted towards automated MOS prediction through end-to-end training of deep neural networks (DNN). Instead of training a network from scratch, we propose to leverage the speech representations from the pre-trained wav2vec-based XLS-R model. However, the number of parameters of such a model exceeds task-specific DNNs by several orders of magnitude, which poses a challenge for resulting fine-tuning procedures on smaller datasets. Therefore, we opt to use pre-trained speech representations from XLS-R in a feature extraction rather than a fine-tuning setting, thereby significantly reducing the number of trainable model parameters.
We compare our proposed XLS-R-based feature extractor to a Mel-frequency cepstral coefficient (MFCC)-based one, and experiment with various combinations of bidirectional long short term memory (Bi-LSTM) and attention pooling feedforward (AttPoolFF) networks trained on the output of the feature extractors. We demonstrate the increased performance of pre-trained XLS-R embeddings in terms a reduced root mean squared error (RMSE) on the ConferencingSpeech 2022 MOS prediction task. 
\end{abstract}


\section{Introduction}

Given the ubiquitous dependence on online conferencing applications in recent years, a reliable and automated assessment of speech quality has become increasingly important. Various factors such as jitter, latency, echo, packet loss, and distortion may affect listening, talking and conversational quality \cite{grancharov_speech_2008}. Ultimately, a reliable assessment of speech quality can help to guide and evaluate speech enhancement approaches (e.g., those presented in the ConferencingSpeech 2021 challenge \cite{rao_interspeech_2021}). Typically, speech quality assessment approaches are grouped into intrusive and non-intrusive methods, depending on whether they operate on the original unprocessed or the processed audio signal. Since objective metrics such as the speech-to-reverberation modulation energy ratio (SRMR) do not necessarily correlate with the perceived speech quality \cite{avila_non-intrusive_2019}, human judgements in the form of Absolute Category Ratings (ACRs) are commonly used in non-intrusive settings.

The resulting ratings are then used to derive mean opinion score (MOS) values as a final subjective assessment of the speech quality of a given audio sample. However, the collection of human judgements is time- and labor-intensive and not feasible for large-scale evaluations of speech quality in online conferencing applications. Furthermore, efforts have been dedicated towards Machine Learning (ML) approaches for the automated prediction of MOS values for given speech fragments \cite{leng_mbnet_2021, tseng_utilizing_2021}, for example based on long short-term memory (LSTM) networks \cite{cauchi_non-intrusive_2019}. More recently, unsupervised pre-trained speech models such as wav2vec 2.0 \cite{baevski_wav2vec_2020} have made it possible to leverage pre-trained representations for specific fine-tuning tasks such as MOS prediction. This is typically realized by updating all model parameters using task-specific data and objective functions. While such models are typically pre-trained on English speech, there are also multilingual variants like \mbox{XLS-R}~\cite{babu_xls-r_2021}, consisting of models with up to two billion parameters trained on 436,000 hours of unlabeled speech.

Nonetheless, fine-tuning such a large number of model parameters is computationally intensive and thus not always an expedient solution. Moreover, while benchmarks (e.g., the combination of multiple automatic speech translation, automatic speech recognition and speech classification tasks that was used to evaluate wav2vec 2.0 \cite{baevski_wav2vec_2020}) aim to assess the generalizability of the performance of fine-tuning pre-trained language or speech representations on downstream tasks, they do not always translate into robustly improved performance on tasks in a specific domain or language \cite{ethayarajh_utility_2020}. As a result, the use of pre-trained models as fixed feature extractors for much smaller task-specific ML architectures might pose a more effective alternative to computationally intensive fine-tuning procedures.

That is why in this paper, we evaluate the potential of using the XLS-R model in a fixed feature extraction rather than a fine-tuning setting for MOS prediction for the ConferencingSpeech 2022 challenge. Our contributions can be summarized as follows. First, in addition to the two baselines provided in the ConferencingSpeech challenge, we propose a Mel-frequency cepstral coefficient (MFCC)-based approach serving as an additional baseline. Secondly, we train a MOS prediction model consisting of a bidirectional LSTM (Bi-LSTM) and an attention pooling feedforward (AttPoolFF) network based on the XLS-R feature extractor. Additionally, we test various combinations of the BiLSTM and AttPoolFF components placed on top of both the XLS-R and MFCC feature extractors. We show that the XLS-R feature extractor does not only outperform the classic MFCC feature extractor, but it also results in competitive model performances with regards to the root mean squared error (RMSE) on data from the ConferencingSpeech 2022 challenge~\cite{yi2022conferencingspeech}.


\section{Methods}

\subsection{Datasets}

The dataset provided in the ConferencingSpeech 2022 challenge comprises four corpora, each with respective MOS labels in a rating range of 1--5, which are derived from ACRs based on the International Telecommunication Union Telecommunication Standardization Sector (ITU-T) recommendation P.808~\cite{naderi_open_2020}. The only exception is the IU~Bloomington corpus, which follows ITU-R BS.1534~\cite{series2014method} and has a range of 0--100.


\subsubsection{Tencent}
The Tencent corpus consists of around 14,000 Chinese speech clips from a total of 1,089 speakers, including clips with and without reverberation. The length of the speech clips ranges from 5s to 13.5s.

\subsubsection{Non-intrusive speech quality assessment (NISQA)}
The NISQA corpus~\cite{mittag_nisqa_2021} is a publicly available collection of more than 14,000 English and German speech clips with real as well as simulated noise. 

\subsubsection{Indiana University (IU) Bloomington}
The IU Bloomington corpus encompasses a total of 36,000 English speech samples from the Voices Obscured in Complex Environmental Settings (VOiCES)~\cite{richey_voices_2018} and COnversational Speech In Noisy Environments (COSINE)~\cite{stupakov_cosine_2009} datasets. All speech clips are between three and six seconds long.

\subsubsection{Public switched telephone network (PSTN)}
The PSTN corpus consists of roughly 80,000 English speech samples based on LibriVox with a length of 10 seconds each, containing both clean samples as well as samples with artificially added background noise.

\subsection{Dataset Division} \label{se:dataset_division}

The original dataset split for training and development provided by the challenge organizers consists of 80\% of the Tencent corpus and 95\% of the PSTN corpus. The remaining 20\% and 5\% of these corpora along with some samples from an additional corpus called TUB compose the test split.

We take two approaches to dataset division. In the first division, we aim to maximize the performance on the challenge's test split. Since the proportion of TUB samples is rather small (only 10\% of the test set), we expect the distribution of the test split to be similar to that of the original training and development split. For this reason, we only include samples from the Tencent and PSTN corpora in this approach.

In the second division, we attempt to compare the generalizability of the various models by training and evaluating the models on a larger pool of datasets. For simplicity, this has been done by shuffling the samples in all corpora (IU~Bloomington, NISQA, PSTN, Tencent) and using 85\% for the training set and 15\% for the validation set.

In order to compare model performance in a correct way, we construct the first division as subsets of the second division's training and validation sets, where only Tencent and PSTN samples are included. Since training is performed on the subsets, there is no overlap between the training samples from one approach and the validation samples from another approach.

\subsection{Preprocessing} \label{se:preprocessing}

The raw audio is processed with a two-step procedure in order to generate the features for the MOS prediction-specific model architectures (see Section \ref{se:models}).

\noindent
\subsubsection{Feature Extraction}
First, feature extraction is performed on the raw audio. For MFCC calculation, we use the implementation by torchaudio \cite{yang_torchaudio_2022} with the default parameters and a sample rate of 16~kHz.
The XLS-R feature extraction is based on the \textit{facebook/wav2vec2-xls-r-300m} model available at the HuggingFace \cite{wolf_transformers_2020} model hub.
Note that no fine-tuning is performed on the feature extraction modules.

\subsubsection{Input Normalization}
After the features are extracted for the entire dataset, the mean and variance of each channel are calculated for the training split. These parameters will be used to normalize the input data.

\subsection{Models}
\label{se:models}

We compare a total of six different model architectures, namely two baselines provided in the ConferencingSpeech 2022 challenge and four novel architectures based on combining different feature extractors and MOS prediction models.\footnote{Our Code: {\scriptsize \url{github.com/lcn-kul/conferencing-speech-2022}}}

\subsubsection{Baseline}

The challenge organizers have provided two baseline models based on the NISQA speech quality prediction model~\cite{mittag_nisqa_2021}.
The overall architecture consists of three parts. First, framewise features are calculated using a feed-forward or convolutional network. Next, the features are updated using a time-dependent network, e.g., a (Bi-)LSTM or a transformer model. Finally, the updated features are pooled and mapped to a one-dimensional output.

The first baseline model (Baseline1) uses a deep feed-forward network followed by an LSTM network and average pooling head. The model was trained for a total of 50 epochs on the entire development set.
The second baseline model (Baseline2) uses a convolutional neural network (CNN) followed by a self-attention network and attention-pooling head. This model was trained on the entire development set for 80 epochs. 

\subsubsection{Proposed Models}

\setlength{\belowcaptionskip}{-10pt}
\begin{figure}[t!]
  \centering
  \includegraphics[width=\linewidth]{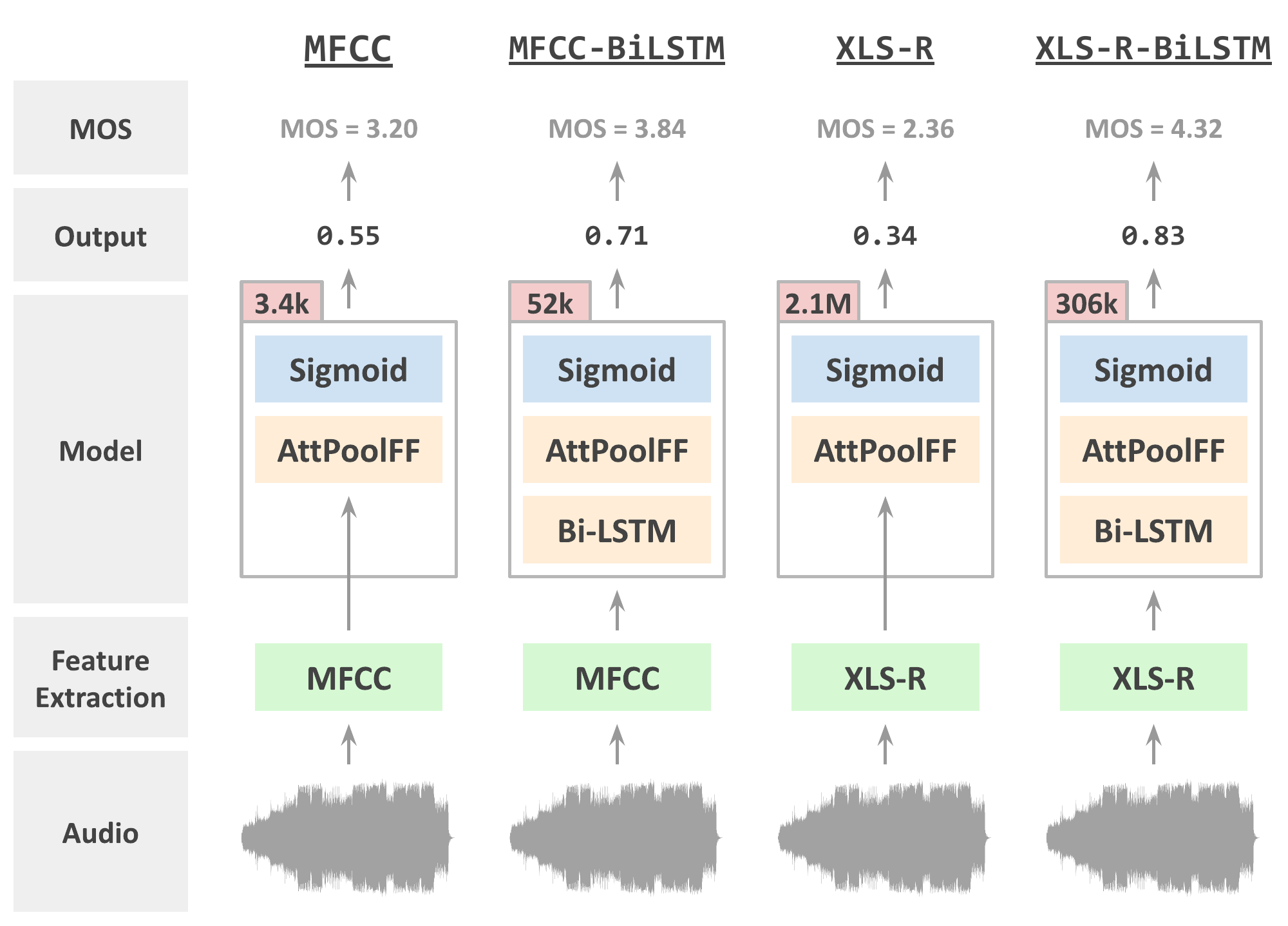}
  \caption{Overview of the model components of all four model variants introduced in this paper. The MFCC and XLS-R networks (see green boxes) are used as fixed feature extractors. The Bi-LSTM and Attention-Pooling layers are trained on the challenge data. The resulting number of trainable parameters for each model are listed in the red boxes.}
  \label{fig:model_architectures}
\end{figure}

The proposed model architectures are illustrated in Figure~\ref{fig:model_architectures}. Each model consists of two parts: a frozen feature extraction module (see Section \ref{se:preprocessing}) and a tunable regression head. We have combined two possible feature extractors (MFCC, XLS-R) with two possible regressors (AttPoolFF, Bi-LSTM + AttPoolFF) which makes four total models.

The goal of these four models is to demonstrate the power of state-of-the-art pre-trained speech embeddings (\mbox{XLS-R}) compared to more traditional features (MFCC) and training embeddings from scratch (baseline). Furthermore, we investigate if additional temporal support through a Bi-LSTM improves performance. 

The regressor input is a sequence of 384 extracted features (roughly 8 seconds of audio). The Bi-LSTM network (if included in the model architecture) uses 2 layers and a size of 32 in each direction.
The attention pooling network with feedforward connection (AttPoolFF) comes from the NISQA model~\cite{mittag_nisqa_2021} and uses a hidden size of 64.  We apply batch normalization before the Bi-LSTM network and dropout of 0.1 after each layer.
Subsequently, the outputs from the AttPoolFF module are mapped to the range (0,1) with a sigmoid function. We refer to these final outputs as normalized MOS values. This intermediate normalization step ensures that the output range of the AttPoolFF layer is unrestricted. During the evaluation of the model, the normalized MOS predictions can be mapped to the original 1--5 MOS range.

\subsection{Evaluation}

Model performance is assessed on a total of three datasets. First, we evaluate our proposed models on our own validation split (see Section~\ref{se:dataset_division}). Secondly, to estimate the performance on the test set, we also evaluate our models on the subset of validation data that come from the Tencent and PSTN corpora. Lastly, we report the test performance of all models on the test data provided by the challenge organizers. We report the RMSE, Pearson correlation coefficient (PCC) and Spearman’s rank correlation coefficient (SRCC) on the aforementioned validation and test datasets. More specifically, these metrics are calculated by first mapping the normalized MOS values (see Section \ref{se:models}) back to the original range of 1--5.

\subsection{Training Details}

The models are implemented using the PyTorch (v.1.11.0)
and PyTorch Lightning (v.1.5.10)
libraries in Python 3.9. Training is performed using the PyTorch Lightning trainer. The network is trained using the ADAM optimizer~\cite{kingma2014adam}, a batch size of 64, and  an MSE loss. We use a cyclical learning rate scheduler~\cite{smith2017cyclical} with a base learning rate of $10^{-3}$ and a maximum learning rate of $10^{-2}$. Each model is trained with a batch size of 64 for a total of 50 epochs, and the model with the lowest validation loss is selected. Each model was trained using an NVIDIA GeForce GTX 1660 Ti graphics card with 6~GB memory.




\section{Results}

\newcommand{\drawtableheader}{
\textbf{} &
\multicolumn{4}{l}{\textbf{Baseline}} &
\multicolumn{4}{l}{\textbf{MFCC}} &
\multicolumn{4}{l}{\textbf{XLS-R}} \\ 
\cmidrule(l){2-13} \textbf{} &
\multicolumn{2}{c}{\multirow{2}{*}{Baseline1}} & 
\multicolumn{2}{c}{\multirow{2}{*}{Baseline2}} & 
\multicolumn{2}{c}{\multirow{2}{*}{MFCC}} & 
\multicolumn{2}{c}{MFCC-} & 
\multicolumn{2}{c}{\multirow{2}{*}{XLS-R}} & 
\multicolumn{2}{c}{XLS-R-} \\ 
\textbf{} &
\multicolumn{2}{c}{\textbf{}} &
\multicolumn{2}{c}{\textbf{}} &
\multicolumn{2}{c}{\textbf{}} &
\multicolumn{2}{c}{Bi-LSTM} &
\multicolumn{2}{c}{\textbf{}} &
\multicolumn{2}{c}{Bi-LSTM} \\ 
\midrule
}

\newcommand{\tableformat}{
@{}lclclclclclcl@{} 
}

\begin{table*}[t!]
\centering
\begin{tabular}{\tableformat}
\toprule
\drawtableheader
\textbf{\begin{tabular}[c]{@{}l@{}}Validation PCC\\ All Corpora\end{tabular}} &
\multicolumn{2}{c}{0.7054} & \multicolumn{2}{c}{0.8182} & \multicolumn{2}{c}{0.5763} & \multicolumn{2}{c}{0.7294} & \multicolumn{2}{c}{0.7835} & \multicolumn{2}{c}{\textbf{0.8398}} \\
\midrule

\textbf{\begin{tabular}[c]{@{}l@{}}Validation PCC\\ PSTN \& Tencent\end{tabular}} & \multicolumn{2}{c}{0.8017} & \multicolumn{2}{c}{\textbf{0.8770}} & \multicolumn{2}{c}{0.7114} & \multicolumn{2}{c}{0.8049} & \multicolumn{2}{c}{0.8058} & \multicolumn{2}{c}{0.8617} \\
\addlinespace[0.5em]


\textbf{\begin{tabular}[c]{@{}l@{}}Test PCC \\ PSTN \& Tencent\end{tabular}} & \multicolumn{2}{c}{0.551} & \multicolumn{2}{c}{0.724} & \multicolumn{2}{c}{/} & \multicolumn{2}{c}{/} & \multicolumn{2}{c}{/} & \multicolumn{2}{c}{\textbf{0.812}} \\ \bottomrule

\addlinespace[2em]
\end{tabular} 

\caption{Pearson correlation coefficient (PCC) of all models on different splits of the datasets provided in the ConferencingSpeech 2022 challenge. For each dataset, the best performance (i.e., the model with the highest PCC) is highlighted in bold.} 
\label{tab:results_pcc}
\end{table*}

\begin{table*}[t!]
\centering
\begin{tabular}{\tableformat}
\toprule
\drawtableheader
\textbf{\begin{tabular}[c]{@{}l@{}}Validation SRCC\\ All Corpora\end{tabular}} &
\multicolumn{2}{c}{0.6995} & \multicolumn{2}{c}{0.8186} & \multicolumn{2}{c}{0.5769} & \multicolumn{2}{c}{0.7236} & \multicolumn{2}{c}{0.7755} & \multicolumn{2}{c}{\textbf{0.8323}} \\ 
\midrule

\textbf{\begin{tabular}[c]{@{}l@{}}Validation SRCC\\ PSTN \& Tencent\end{tabular}} & \multicolumn{2}{c}{0.7954} & \multicolumn{2}{c}{\textbf{0.8712}} & \multicolumn{2}{c}{0.7068} & \multicolumn{2}{c}{0.7986} & \multicolumn{2}{c}{0.7976} & \multicolumn{2}{c}{0.8543} \\ 
\addlinespace[0.5em]

\textbf{\begin{tabular}[c]{@{}l@{}}Test SRCC \\ PSTN \& Tencent\end{tabular}} & \multicolumn{2}{c}{/} & \multicolumn{2}{c}{/} & \multicolumn{2}{c}{/} & \multicolumn{2}{c}{/} & \multicolumn{2}{c}{/} & \multicolumn{2}{c}{/} \\ \bottomrule 
\addlinespace[2em]
\end{tabular}

\caption{Spearman’s rank correlation coefficient (SRCC) of all models on different splits of the datasets provided in the ConferencingSpeech 2022 challenge. For each dataset, the best performance (i.e., the model with the highest SRCC) is highlighted in bold. At the time of writing, no SRCC test results have been made public.} 
\label{tab:results_srcc}
\end{table*}

\begin{table*}[t!]
\centering
\begin{tabular}{\tableformat}
\toprule
\drawtableheader

\textbf{\begin{tabular}[c]{@{}l@{}}Validation RMSE\\ All Corpora\end{tabular}} &
\multicolumn{2}{c}{0.8224} & \multicolumn{2}{c}{0.5475} & \multicolumn{2}{c}{0.7579} & \multicolumn{2}{c}{0.6227} & \multicolumn{2}{c}{0.5674} & \multicolumn{2}{c}{\textbf{0.5000}} \\ 
\midrule

\textbf{\begin{tabular}[c]{@{}l@{}}Validation RMSE\\ PSTN \& Tencent\end{tabular}} & \multicolumn{2}{c}{0.6614} & \multicolumn{2}{c}{0.4965} & \multicolumn{2}{c}{0.6585} & \multicolumn{2}{c}{0.5551} & \multicolumn{2}{c}{0.5543} & \multicolumn{2}{c}{\textbf{0.4759}} \\ 
\addlinespace[0.5em]


\textbf{\begin{tabular}[c]{@{}l@{}}Test RMSE \\ PSTN \& Tencent\end{tabular}} & \multicolumn{2}{c}{0.745} & \multicolumn{2}{c}{0.543} & \multicolumn{2}{c}{/} & \multicolumn{2}{c}{/} & \multicolumn{2}{c}{/} & \multicolumn{2}{c}{\textbf{0.344}} \\ \bottomrule 

\addlinespace[2em]
\end{tabular} 

\caption{Root mean squared error (RMSE) of all models on different splits of the datasets provided in the ConferencingSpeech 2022 challenge. For each dataset, the best performance (i.e., the model with the lowest RMSE) is highlighted in bold.} 
\label{tab:results_rmse}
\end{table*}

\renewcommand{\arraystretch}{1.4}
\begin{table*}[t!]
\centering
\begin{tabular}{|c|ccc|ccc|ccc|ccc|}
\hline

&
\multicolumn{3}{|c|}{Entire Test Set} &
\multicolumn{3}{|c|}{Only PSTN} &
\multicolumn{3}{|c|}{Only TUB} &
\multicolumn{3}{|c|}{Only Tencent} \\
\hline

Model & 
PCC & RMSE & RMSE-S &
PCC & RMSE & RMSE-S &
PCC & RMSE & RMSE-S &
PCC & RMSE & RMSE-S \\

\hline

Baseline1 & 
0.530 & 0.768 & 0.497 &
0.361 & 0.585 & 0.293 &
0.348 & 1.094 & 0.649 &
0.881 & 0.624 & 0.550 \\

\textbf{Ours} & 
\textbf{0.800} & \textbf{0.365} & \textbf{0.320} &
\textbf{0.636} & \textbf{0.351} & \textbf{0.247} &
\textbf{0.796} & \textbf{0.443} & \textbf{0.418} &
\textbf{0.967} & \textbf{0.300} & \textbf{0.295} \\

\hline

\addlinespace[2em]
\end{tabular} 

\caption{Preliminary test results for Baseline1 and our submitted XLS-R Bi-LSTM model. RMSE-S is the RMSE score smoothed by a third-order polynomial function. Our model outperforms Baseline1 on all metrics. The values differ slightly from those of Tables~\ref{tab:results_pcc} and~\ref{tab:results_rmse}. This is because extra test samples were added to the TUB dataset after the preliminary results were announced.}
\label{tab:results_published}
\end{table*}

\subsection{Validation Results}
\label{se:validation_results}
Table~\ref{tab:results_rmse} shows that the XLS-R Bi-LSTM model that was trained on the Tencent/PSTN samples outperforms the baseline with a RMSE reduction of 0.0206 (4.1\% improvement) on the respective validation set. It is noteworthy that the baseline's training set included these samples, so it would have a slight advantage in this comparison. The MFCC-based models are substantially worse than the baseline model.

Surprisingly, the XLS-R model without a Bi-LSTM module performs much worse than its Bi-LSTM counterpart.
We investigated this result by training the same model on the larger split including all four corpora. This resulted in rapid convergence and better performance (PCC: 0.8426, SRCC: 0.8372, RMSE: 0.5042 on validation split of reduced dataset). We therefore conclude that the low performance is likely because the reduced dataset is too small for the number of model parameters.

For the PCC and SRCC metrics, shown in Tables~\ref{tab:results_pcc} and~\ref{tab:results_srcc} respectively, the baseline still outperforms our best model by 0.0153 (1.8\% better) and 0.0169 (2.0\% better) respectively.
Further investigation is required to determine the reason why the XLS-R-based models perform similar to or better than the baseline on RMSE but worse for PCC and SRCC.

\subsection{Test Results}
\label{se:test_results}



The final row of Tables~\ref{tab:results_pcc} and~\ref{tab:results_rmse} shows the challenge results for the PCC and RMSE metrics respectively. Our XLS-R Bi-LSTM model outperforms the baselines for both metrics. For the PCC metric this is an improvement of 0.261 (47.4\% better) over Baseline1 and 0.088 (12.2\% better) over Baseline2. For the RMSE metric, our model shows a reduction of 0.401 (53.8\% better) and 0.199 (36.6\% better) over the baselines.

Our model also outperforms the other 17 competitors.\footnote{Results: {\scriptsize \url{tea-lab.qq.com/challenge_rankings_2022.pdf}}} Compared to the next best model, our PCC and RMSE metrics improve by 0.015 (1.9\% better) and 0.092 (21.1\% better) respectively. Finally, for the smoothed RMSE metric, which is the primary evaluation metric, we improve by 3.9\% compared to the next best model and by 37.3\% and 20.3\% compared to the baselines.

The large gap in RMSE metrics is due to outstanding performance on the unseen TUB dataset. For the PSTN and Tencent corpora, our model performs comparably to the next best model (1.7\% better and 5.6\% worse). But for the TUB corpus, our model performed 27.4\% better than the next-closest competitor for that corpus and 42.9\% better than the overall second-place model~\cite{yi2022conferencingspeech}. This suggests that the proposed model performs exceptionally well on unseen data. Further research is required to validate this claim. 

\section{Conclusions}

In this paper, we have presented our submission for the ConferencingSpeech 2022. Our model outperforms the baselines by 53.8\% and 36.6\% and also the next best competitor by 21.1\% on the RMSE metric. Through this work, we are able to demonstrate the effectiveness of using pre-trained wav2vec-based XLS-R in a feature extraction setting.

In future work, we aim to compare the generalizability of the wav2vec-based models versus the baseline model by evaluating them on larger unseen datasets. Finally, we would investigate the use of a Transformer instead of a Bi-LSTM in the model architectures. 

\section{Acknowledgements}

This work was supported by KU Leuven Bijzonder Onderzoeksfonds (BOF) under grant C14/21/109. 


\bibliographystyle{IEEEtran}
\bibliography{mybib}

\end{document}